# Retrieving the Complex Intracavity Pump Field of a Kerr Comb from the Through Port Data

Xiaoxiao Xue[1*], Yi Xuan[1,2], Yang Liu[1], Pei-Hsun Wang[1], Steven Chen[1], Jian Wang[1,2], Dan E. Leaird[1], Minghao Qi[1,2], and Andrew M. Weiner[1,2]

[1]School of Electrical and Computer Engineering, Purdue University, 465 Northwestern Avenue, West Lafayette, Indiana 47907-2035, USA
[2]Birck Nanotechnology Center, Purdue University, 1205 West State Street, West Lafayette, Indiana 47907, USA
*xue40@purdue.edu

**Abstract:** A method of retrieving the complex intracavity pump field from the through port is proposed, and verified through characterizing the time-domain waveform of a mode-locked comb related to dark soliton formation in a normal-dispersion microresonator.

Optical frequency combs generated from microresonators with a high nonlinear Kerr effect show great potential to achieve ultra-compact size, high repetition rate, and large spectral width [1]-[5]. From the frequency domain perspective, the broadband comb lines grow up due to modulational instability gain and cascaded four-wave mixing. Time-domain investigations of how the intracavity field evolves provide further insights into the coherence of comb lines and the mode-locking mechanism [6]-[10]. However, it is not a straight-forward task to probe the intracavity field from the through port of the bus waveguide which is generally used for both pump injection and internal comb extraction, because the pump line at the through port is the coherent summation of two fields: the input field that passes by the microresonator and the field coupled out of the microresonator. Using another waveguide coupled to the microresonator which acts as a drop port can solve this problem [11], but at the expense of increased complexity and total cavity loss. Here we propose a method of retrieving the intracavity complex pump based on through port measurements. The increased loss seen by the pump due to power transfer to the generated comb as well as the cold-cavity linear loss is considered to determine the effective pump-cavity detuning. The complex intracavity pump is then calculated based on the cavity build-up equation. The intracavity field is successfully reconstructed based on the through port data and verified through comparison to the result obtained from the drop port.

In comb generation, the effective cavity loss for the pump line (denoted by $\alpha_{\text{eff}}$ which takes into account the power transfer to the other comb lines) can be calculated according to the cold-cavity loss and the comb power distribution internal to the resonator. The amplitude drop of the pump line in comb generation compared to when the pump frequency is out of resonance can be measured in experiments and is given by

$$|L| = \left| \left[ -\theta/(\alpha_{\text{eff}} + i\delta_{\text{eff}}) + \sqrt{1-\theta} \right] \Big/ \sqrt{1-\theta} \right| \qquad (1)$$

The correction factor, defined as the ratio of the complex pump field coupled out of the resonator over the coherent summation value at the through port, is given by

$$C = \left[ -\theta/(\alpha_{\text{eff}} + i\delta_{\text{eff}}) \right] \Big/ \left[ -\theta/(\alpha_{\text{eff}} + i\delta_{\text{eff}}) + \sqrt{1-\theta} \right] \qquad (2)$$

where $\theta$ is the coupling intensity loss; $\delta_{\text{eff}}$ is the effective pump-cavity phase detuning. Equations (1) and (2) are two independent equations for $\delta_{\text{eff}}$ and $|C|$, and can be solved numerically. With the obtained correction factor $C$, the intracavity field can be reconstructed based on the corrected pump and the other comb lines.

The experiments are performed with a silicon nitride microring resonator. The microscope image of the microring is shown in Fig. 1. The radius is 100 μm corresponding to a free spectral range (FSR) of ~231 GHz. The cross-section dimension of the ring waveguide is 2 μm X 550 nm corresponding to a normal dispersion of ~−150 ps/nm/km. Besides the through waveguide, a drop waveguide is also fabricated. The intracavity field is also characterized at the drop port as a reference for checking our method. The coupling gap between the through/drop waveguide and the microring is 500 nm. Figure 2 shows the measured through-port transmission of the resonance which is pumped for comb generation. The loaded quality factor is $8.6 \times 10^5$. The power transfer to the drop port is included as part of the cavity loss. The phase curve indicates that the ring is under-coupled. Figure 3 shows the experimental setup of measuring the phase of each comb line through line-by-line shaping in combination with autocorrelation measurement [6]. The fiber dispersion between the through/drop port and the input of autocorrelator is roughly compensated by using a length of dispersion-compensating fiber (DCF) and finely compensated by programming the pulse shaper. The insertion loss of the microring chip, including the fiber coupling loss and the bus waveguide transmission loss, is 4.4 dB. The pump power in front of the chip is 0.8 W. Figures 4(a) and 4(b) show the comb spectra measured at the through port and the drop port, respectively. The comb is initialized by mode interaction taking place 2-FSR away from the pump to the shorter wavelength [12]-[13]. The comb line spacing is 2 FSRs. By programming the pule shaper, the time-domain waveform can be cleanly compressed to a transform-limited pulse (autocorrelation trace not shown), which implies a high coherence between the comb lines. Since the fiber link dispersion is compensated, shaping the comb to a transform-limited pulse allows us to probe the original phase of each comb line right after the microring chip. The line-by-line shaping procedure is performed at both the through port and the drop port. Figure 5(a) and 5(b) show the measured comb intensity and phase respectively. As can be expected, the complex comb amplitude at the through port and drop port are very close except for the pump line. We then use the procedure outlined above to correct the pump line as measured at the through port to estimate its complex amplitude internal to the ring. The corrected pump amplitude is compared to the value obtained from the drop port measurement, which should be directly proportional to the internal field without correction. As shown in Fig 5, the corrected complex pump amplitude now gets very close to the drop port value. The intensity agreement is improved from 4.25 dB to 0.58 dB; and the phase agreement is improved from −2.26 rad to 0.14 rad. Figures 6(a) and 6(b) show the reconstructed intracavity field based on the comb intensity and phase measured at the through port and the drop port, respectively. The result retrieved from the through port is very close to that from the drop port which verifies the validity of our method. The time-domain intracavity field takes the form of dark solitons [14] which are softly excited from low-level noise with the aid of mode interaction [10]. In conclusion, a method of retrieving the intracavity complex pump from the through port is proposed, and verified through characterizing the time-domain waveform of dark solitons in a normal-dispersion microresonator.

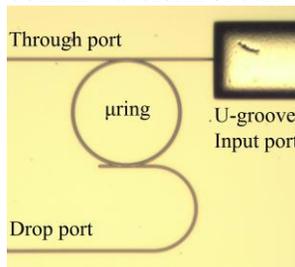

Fig. 1. Microscope image of the SiN microring.

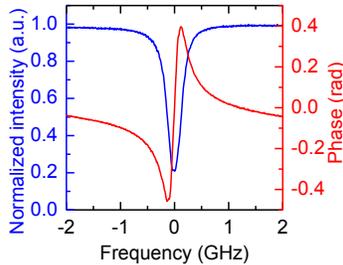

Fig. 2. Measured intensity and phase of the resonance excited for comb generation.

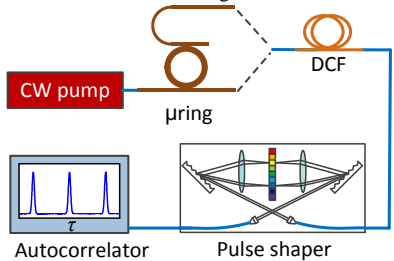

Fig. 3. Experimental setup of Kerr comb generation and line-by-line shaping.

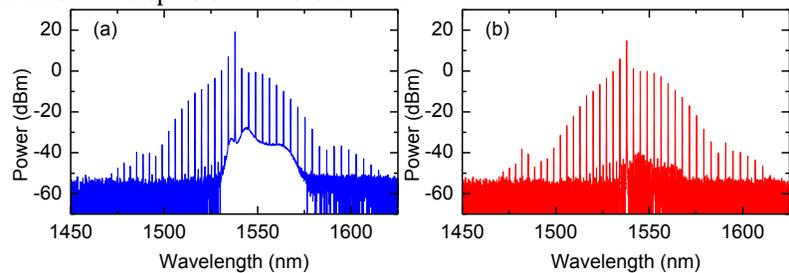

Fig. 4. Comb spectra measured at the (a) through port and the (b) drop port.

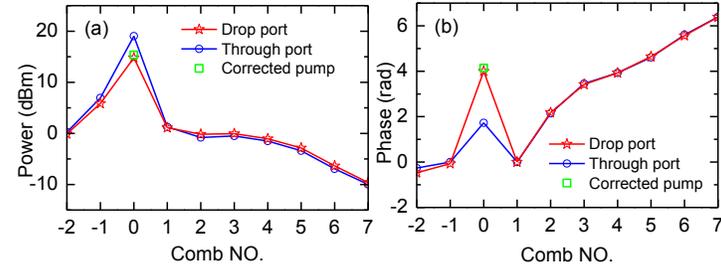

Fig. 5. Retrieved intensity (a) and phase (b) of each comb line at the through port and the drop port. Also shown are the corrected pump intensity and phase at the through port.

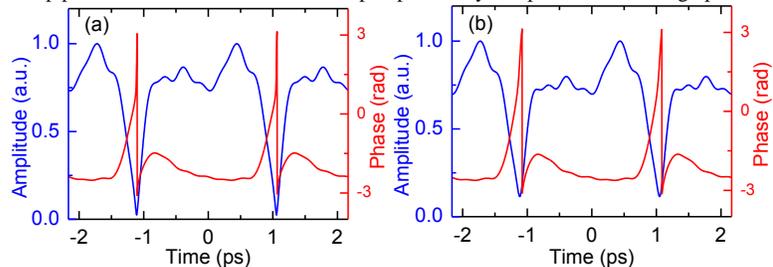

Fig. 6. Reconstructed intracavity field based on (a) corrected through port data and (b) drop port data.